\documentclass[pra,twocolumn,floatfix,aps]{revtex4}

\usepackage{gensymb}
\usepackage{amsmath}
\usepackage{makeidx}
\usepackage{amssymb}
\usepackage{amsmath}
\usepackage{graphicx}
\usepackage{ulem}
  
\usepackage[usenames]{color}
\usepackage{hyperref}  
\usepackage[T1]{fontenc}    

\begin{document}

\title{Spontaneous dipolar Bose-Einstein condensation  on the surface of a cylinder}

\author{Luis E. Young-S.$^{1}$\footnote{lyoung@unicartagena.edu.co}}

 \author{S. K. Adhikari$^2$\footnote{sk.adhikari@unesp.br,  professores.ift.unesp.br/sk.adhikari/}}
 
 \affiliation{$^1$Grupo de Modelado Computacional y Programa de Matem\'aticas, Facultad de Ciencias Exactas y Naturales, 
 Universidad de Cartagena, 130015 Cartagena de Indias, Bolivar, Colombia}

%\affiliation{$^2$Instituto de Matemáticas Aplicadas, Universidad de Cartagena,  130001 Cartagena de Indias, Bolivar, Colombia}

\affiliation{$^2$Instituto de F\'{\i}sica Te\'orica, UNESP - Universidade Estadual Paulista, 01.140-070 S\~ao Paulo, S\~ao Paulo, Brazil}

\begin{abstract}

We demonstrate the spontaneous formation of  a Bose-Einstein condensate (BEC)   of strongly-bound 
harmonically-trapped dipolar $^{164}$Dy  atoms  on the outer curved surface of an  elliptical   or a circular cylinder, with a distinct topology, employing the numerical solution of  an improved  mean-field model including a   Lee-Huang-Yang-type interaction,  meant to stop a collapse at high atom density, the axis of the   cylindrical-shell-shaped BEC being aligned along the polarization direction of the dipolar atoms. These states are dynamically stable and a Gaussian  initial state leads to the  cylindrical-shell-shaped state in both imaginary-time and real-time propagation.  The formation of the hollow cylindrical BEC by a real-time simulation starting from a solid cylindrical state demonstrate the possibility of the formation of such a condensate experimentally.

\end{abstract}

% \pacs{03.75.Hh, 03.75.Mn, 03.75.Kk, 03.75.Lm}

\maketitle

\section{Introduction}

After the experimental observation of a Bose-Einstein condensate (BEC) of alkali-metal atoms,
this system has been the laboratory testing ground of different quantum phenomena, not accessible 
for experimental study previously.  However, practically all of these studies were confined to a normal three-dimensional (3D) space or a two-dimensional (2D) plane or a one-dimensional straight line. Nevertheless, 
new physics may appear in curved  spaces with distinct topology. 
Quantum states with distinct topology  \cite{top1,top2,top3,curved} have been the subject matter of intense investigation in recent times  not only in search of new physics but also due to its possible application in  
quantum information processing \cite{QC}. The nature of vortices on curved surfaces is in general different 
from the same in 3D space \cite{vor}.
New  types of vortices may appear in spinor condensates on a curved surface with distinct topology in synthetic gauge field \cite{HH,HH1}.
Fractional quantum Hall   states possess a richer structure evident through their response to
topology \cite{FQH}. Unique  superfluid properties have been pointed out to exist in a BEC in a ring trap \cite{ring1,ring2}.

 Consequently, 
in recent years there has been a great deal of interest in the investigation  and the observation of a BEC on a curved surface   with distinct topology
and 
experimental studies have begun to 
investigate properties of a BEC formed on a spherical surface in the form of a spherical bubble.
  This is a difficult task in the presence of gravity, which will bring down the atoms from the top of the bubble.   
To circumvent this difficulty, { there has been an attempt to create}
a
single-species BEC of $^{87}$Rb atoms   on a spherical bubble 
 in  orbital microgravity \cite{BBL-space} in a space-based quantum gas laboratory \cite{lab,lab2,EO}
 following  a suggestion  \cite{sugg1,sugg2} and elaborations in Ref.  \cite{TS,sugg3,sugg4,sugg5}.  
Optically trapped spherical-shell-shaped  binary BEC of $^{23}$Na and $^{87}$Rb
atoms in the immiscible phase has been created and studied in a laboratory on earth \cite{SHL-erth}, in the presence of gravity, following a suggestion  \cite{Ho,Pu}  and consideration by  others \cite{bin1,bin2,bin3}.
{Hemispherical}-shell-shaped BEC of $^{87}$Rb atoms   has also been created in a laboratory on earth \cite{bs2}  by  
a novel gravity compensation mechanism. 
{All these    experiments \cite{BBL-space,SHL-erth,bs2} for  the creation of  a  spherical-shell-shaped BEC  require a complex  engineering of the  external (confining) parameters. 
The same is true for a cylindrical-shell-shaped or a ring-shaped BEC created in a laboratory \cite{ring1,ring2}.}
{It is highly desirable to have   a shell-shaped BEC in a {\it harmonic} trap with angular frequencies achievable in a laboratory. }

In this paper we demonstrate the formation of a  cylindrical-shell-shaped  BEC of strongly dipolar $^{164}$Dy atoms, {with  distinct topology in the form of a hollow cylinder},
 in a 
strong harmonic trap using    {an improved mean-field}
 model including a Lee-Huang-Yang-type (LHY) interaction \cite{lhy}, appropriately modified for dipolar atoms \cite{qf1,qf3,qf2}, the axis of the cylinder being aligned along the polarization direction of dipolar atoms.  Although a cylindrical shell is topologically equivalent to a ring, it has a different geometrical shape with an extended curved surface.  This will allow study of new physics, like emergence of vortex and vortex lattice on a different curved surface.
  In the framework of the mean-field  Gross-Pitaevskii (GP) equation,
a dipolar BEC  shows collapse instability as the  dipolar interaction increases beyond a critical value \cite{ex1,ex2}, 
and the inclusion of 
 a LHY   interaction \cite{qf1,qf2}
in theoretical investigations     stabilizes the  dipolar
condensate  against collapse \cite{santos}. As the number of atoms $N$ and/or the dipole moment of the atoms 
in a trapped quasi-2D  dipolar BEC  increases,     the condensate  cannot  collapse due to the  repulsive LHY interaction and, 
a droplet \cite{ex1,ex2}, or a spatially-periodic  \cite{ex3}  triangular-, square- \cite{th4},   or  honeycomb-lattice \cite{th4} structure of droplets   or a  labyrinthine state  \cite{pfau,science} could be formed  in a strongly dipolar quasi-2D BEC.  In a strong quasi-2D trap the formation of multiple droplets on a triangular lattice was confirmed in experiments \cite{ex1,ex2,ex3}    and substantiated in theoretical studies \cite{th1,th2,th3,th4,th5}. As the angular frequencies of the trap  are further 
increased, the present  cylindrical-shell-shaped BEC is formed as  the number of atoms is increased beyond  a critical value, where the dipolar atoms are deposited on the outer surface of a hollow cylinder. 
 
 The number of atoms, scattering length and the trap parameters for the formation of a cylindrical-shell-shaped dipolar BEC of $^{164}$Dy atoms 
 are within the reach of ongoing experiments  \cite{expt}.  The scattering length $a$ is taken in the range $85a_0> a>80a_0$, where $a_0$ is the Bohr radius.
  The frequency $f_z$ of the harmonic trap in the polarization $z$ direction is taken in the range $ 250$ Hz  $>f_z > 150$ Hz, whereas the  frequencies $f_x$ and $f_y$  of the harmonic trap in  the transverse $x$-$y$ plane are taken as  $f_x, f_y \approx  0.75 f_z$.    We base our study on a numerical solution of an improved  mean-field model, including the LHY interaction,
  by imaginary-time propagation.   
The number of atoms $N$ is taken in the range $300000>N>100000$; in this study we take it to be mostly
around   $N\approx 200000$, as this number can  easily be achieved in experiments with  $^{164}$Dy atoms \cite{expt,arch}.  For this choice of parameters, the only possible axially-symmetric
state is the  cylindrical-shell-shaped state. %{However, for large $f_z$ ($\gtrapprox 210$ Hz) and $N$ ($\gtrapprox 320000$), we found a cylindrical state with four holes with a different topology as the ground state. The hollow cylindrical state with one hole becomes an excited state in this region.}
 Nevertheless,  for $N\approx 200000$ and  $ 150$ Hz $< f_z \lessapprox 250$ Hz, all Gaussian-type  initial states
 lead to the   cylindrical-shell-shaped state,  in both imaginary- and real-time propagation.     For $N  \lessapprox 100000$, single or multiple  droplet states are preferentially formed. 
We find that for $f_z \lessapprox   120$ Hz, it is difficult to find a cylindrical-shell-shaped state numerically.     As  $f_z$ is increased beyond  $ 250$ Hz
the internal radius of the shell-shaped cylinder gradually decreases and eventually the cylindrical shell may become a solid cylinder and will not be discussed in this study. 
{Moreover, we find that for large $f_z$  ($f_z\gtrapprox 210$ Hz) and large $N$ ($N\gtrapprox 330000$), one could have a  cylindrical state with few (2 to 4) holes aligned parallel to the polarization direction (not studied here).}
If the ratio  $f_x/f_z$  is increased beyond 0.8,  the internal radius of the 
cylindrical-shell-shaped state becomes smaller.  However, as the ratio     $f_x/f_z $ 
 is decreased beyond 0.7, although the internal radius of the cylindrical-shell-shaped state increases,     it  is not the ground state; droplet and multiple-droplet states naturally appear as the ground states  as confirmed in previous  experimental \cite{ex1,ex2,ex3} and theoretical \cite{th1,th2,th3,th4,th5} studies.  
This is why in  this study of cylindrical-shell-shaped states we will mostly consider $N\approx 200000, 150$ Hz $<f_z \lessapprox 250$ Hz, $f_x,f_y =0.75f_z$ and $85a_0>a>80a_0$. 
If the confining harmonic trap  in the $x$-$y$ plane is made slightly anisotropic ($f_x\ne f_y$),  the  cylindrical shell of the 
strongly dipolar BEC becomes elliptical in nature.  

For the experimental value of scattering length $a=92a_0$ \cite{scatmes} of $^{164}$Dy atoms, 
the atomic contact repulsion dominates over the dipolar interaction and 
the dipolar BEC  has the form of a solid cylinder. As the scattering length is gradually reduced, the cylindrical-shell-shaped states appear.  Starting from $a=92a_0$, we demonstrate by real-time propagation that, as the scattering length is gradually reduced by a quasi-linear ramp, 
the cylindrical-shell-shaped state appears for $a=85a_0$  and the shell-shaped structure becomes more pronounced as the scattering length is reduced to $a=80a_0$.  This indicates the viability of the creation of a cylindrical-shell-shaped dipolar BEC in a laboratory by controlling the scattering length near a Feshbach resonance.

In Sec. II we present the improved mean-field model including the LHY interaction \cite{lhy} appropriately modified \cite{qf1,qf2} for dipolar atoms with repulsive atomic interaction (positive scattering length). In Sec. III we present our numerical results for the formation of stationary cylindrical-shell-shaped dipolar BECs obtained by imaginary-time propagation. We also demonstrate the formation the cylindrical-shell-shaped states by real-time propagation starting from a cylindrically-symmetric Gaussian initial state. 
 In Sec. IV we present a brief summary of our findings.

\section{Improved mean-field model}

 We consider a dipolar BEC of $N$ atoms,  polarized along the $z$ direction, of mass $m$ each, and with atomic scattering length $a$. 
The formation of a  cylindrical-shell-shaped BEC   can be formulated  by the following  3D 
GP equation with the inclusion of  the  LHY interaction \cite{dip,dipbec,blakie,yuka}
\begin{align}\label{eq.GP3d}
 \mbox i \hbar \frac{\partial \psi({\bf r},t)}{\partial t} &=\
{\Big [}  -\frac{\hbar^2}{2m}\nabla^2
+U({\bf r})
+ \frac{4\pi \hbar^2}{m}{a} N \vert \psi({\bf r},t) \vert^2 \nonumber\\
& +\frac{3\hbar^2}{m}a_{\mathrm{dd}}  N
\int U_{\mathrm{dd}}({\bf R})
%\frac{1-3\cos^2 \theta}{|{\bf R}|^3}
\vert\psi({\mathbf r'},t)\vert^2 d{\mathbf r}'  \nonumber\\
& +\frac{\gamma_{\mathrm{LHY}}\hbar^2}{m}N^{3/2}
|\psi({\mathbf r},t)|^3
\Big] \psi({\bf r},t), 
\\
%U_{\mathrm{ext}}
%&=\ ,
U({\bf r})&=\frac{1}{2}m(\omega_x^2x^2+\omega_y^2y^2+\omega_z ^2z^2) ,
\\
U_{\mathrm{dd}}({\bf R}) &= \frac{1-3\cos^2 \theta}{|{\bf  R}|^3},\quad a_{\mathrm{dd}}=\frac{\mu_0 \mu^2 m }{ 12\pi \hbar ^2},\label{pd}
\end{align}
where % the confining harmonic-oscillator trap $U(\bf r)$ is given by 
 %$U({\bf r})=\frac{1}{2}m(\omega_x^2x^2+\omega_y^2y^2+\omega_z ^2z^2) $ with angular frequencies 
 $\omega_x (\equiv 2\pi f_x), \omega_y (\equiv 2\pi f_y), \omega_z (\equiv 2\pi f_z)$  are the angular frequencies of the harmonic trap 
 along $x,y,z$ directions, respectively, 
 $\mu_0$ is the permeability of vacuum, $\mu$ is the magnetic dipole moment of each atom, 
 $U_{\mathrm{dd}}({\bf R})$  is the anisotropic dipolar interaction between two atoms located at $\bf r \equiv \{x,y,z\}$ and $\bf r' \equiv \{x',y',z 
'\}$  
and $\theta$ is the angle made by  $\bf R\equiv r-r'$ with the  polarization
$z$ direction, and the dipolar length $a_{\mathrm{dd}}$ measures the strength of atomic dipolar interaction in the same way as the scattering length $a$ measures the strength of atomic contact interaction;
the wave function is  normalized as $\int \vert \psi({\bf r},t) \vert^2 d{\bf r}=1.$

The    LHY interaction  coefficient  $\gamma_{\mathrm{LHY}}$ in  Eq. (\ref{eq.GP3d}) is given by \cite{qf1,qf3,qf2}
\begin{align}\label{qf}
\gamma_{\mathrm{LHY}}= \frac{128}{3}\sqrt{\pi a^5} Q_5(\varepsilon_{\mathrm{dd}}), \quad \varepsilon_{\mathrm{dd}}= \frac{ a_{\mathrm{dd}}}{a},
\end{align}
where  the auxiliary function $ Q_5(\varepsilon_{\mathrm{dd}})$ includes the perturbative quantum-fluctuation
correction due to dipolar interaction \cite{qf1,qf2}  and is
 given by 
\begin{equation}
 Q_5(\varepsilon_{\mathrm{dd}})=\ \int_0^1 dx(1-\varepsilon_{\mathrm{dd}}+3x^2\varepsilon_{\mathrm{dd}})^{5/2},  
\end{equation}
{which can be approximated  as \cite{th1,blakie}
\begin{align}\label{exa}
Q_5(\varepsilon_{\mathrm{dd}}) &=\
\frac{(3\varepsilon_{\mathrm{dd}})^{5/2}}{48}   \Re \left[(8+26\eta+33\eta^2)\sqrt{1+\eta}\right.\nonumber\\
& + \left.
\ 15\eta^3 \mathrm{ln} \left( \frac{1+\sqrt{1+\eta}}{\sqrt{\eta}}\right)  \right], \quad  \eta = \frac{1-\varepsilon_{\mathrm{dd}}}{3\varepsilon_{\mathrm{dd}}},
%\\&\approx 1+ \frac{3}{2}\varepsilon_{\mathrm{dd}}^2
\end{align}
where $\Re$ denotes the real part. Limiting values are to be taken in Eq. (\ref{exa}), for   $\varepsilon_{\mathrm{dd}}=0$ and 1, while $Q_5(0)=1$ and $Q_5(1)= 3\sqrt 3/2$ \cite{th1}.
In this study we use the analytic expression (\ref{exa}) for $Q_5(\varepsilon_{\mathrm{dd}})$.
The dimensionless ratio $
\varepsilon_{\mathrm{dd}}  $   %\equiv {a_{\mathrm{dd}}}/{a}$
determines
the strength of the dipolar interaction relative to  the contact interaction 
and is useful to study  many properties of a dipolar BEC.  The perturbative result (\ref{qf}) is technically valid for 
$\varepsilon_{\mathrm{dd}}<1$, while $Q_5(\varepsilon_{\mathrm{dd}})$ is real \cite{qf1,qf2}. In the domain of present study of strongly dipolar atoms   ($\varepsilon_{\mathrm{dd}}>1$), $Q_5(\varepsilon_{\mathrm{dd}})$ is complex. However, its imaginary part is negligible in  comparison with its real part  in the case of  $^{164}$Dy atoms   \cite{young}   and will be neglected in this study as in  other investigations \cite{th1,th4,th5,expt}.}

Equation (\ref{eq.GP3d}) can be written  in  
the following  dimensionless form if we  scale lengths in terms of   $l = \sqrt{\hbar/m\omega_z}$, time in units of $\omega_z^{-1}$,  angular frequency in units of $\omega_z$,  energy in units of $\hbar\omega_z$
and density $|\psi|^2$ in units of $l^{-3}$
\begin{align}\label{GP3d2}
\mbox i \frac{\partial \psi({\bf r},t)}{\partial t} & =
{\Big [}  -\frac{1}{2}\nabla^2
+{\frac{1}{2}}\left({f_x^2}x^2+ {f_y^2}y^2+z^2\right)
\nonumber\\ &+ 4\pi{a} N \vert \psi({\bf r},t) \vert^2
+3a_{\mathrm{dd}}  N 
\int % \frac{1-3\cos^2 \theta}{|{\bf R}|^3}
U_{\mathrm{dd}}({\bf R})
\vert\psi({\mathbf r'},t)\vert^2 d{\mathbf r}'   \nonumber \\ 
% \end{align} %\nonumber \\ &
%\begin{align}
&+\gamma_{\mathrm{LHY}}N^{3/2}
|\psi({\mathbf r},t)|^3
\Big] \psi({\bf r},t), 
\end{align} 
where all variables are  scaled. We are using the same symbols to represent the scaled  variables as the unscaled ones without any risk of confusion. 
%we are representing both the scaled and unscaled variables by the same symbols. 
 Equation (\ref{GP3d2})  
can be derived  using  the variational principle 
\begin{align}
\mbox i \frac{\partial \psi}{\partial t} = \frac{\delta E}{\delta \psi^*}, %, \nonumber \\
\end{align}
%={\Big [}  -\frac{1}{2}\nabla^2
%+U_{\mathrm{ext}}({\bf r})
%+\frac{1}{2}\Big(\frac{\omega_x^2}{\omega_z^2}x^2+ \frac{\omega_y^2}{\omega_z^2}y^2+z^2\Big)
%\nonumber\\ &+3a_{\mathrm{dd}}  N
%\int \frac{1-3\cos^2 \theta}{|{\bf R}|^3}
%\vert\psi({\mathbf r'})\vert^2 d{\mathbf r}'  \nonumber \\ &+ 4\pi{a} N \vert \psi({\bf r}) \vert^2
%+\gamma_{\mathrm{QF}}N^{3/2}
%|\psi({\mathbf r})|^3
%\Big] \psi({\bf r}).
%\end{align}
which leads to  the following energy functional corresponding to  energy per atom 
\begin{align}
E &= \frac{1}{2}\int d{\bf r} \Big[ {|\nabla\psi({\bf r})|^2} +\Big({f_x^2}x^2+ {f_y^2}y^2+z^2\Big)|\psi({\bf r})|^2\nonumber  \\
% \end{align} \begin{align}
&+ {3}a_{\mathrm{dd}}N|\psi({\bf r})|^2 
\left. \int U_{\mathrm{dd}}({\bf R} )
%\frac{1-3\cos^2\theta}{R^3}
|\psi({\bf r'})|^2 d {\bf r'} \right. \nonumber \\
& + 4\pi Na |\psi({\bf r})|^4 +\frac{4\gamma_{\mathrm{LHY}}}{5} N^{3/2}
|\psi({\bf r})|^5\Big]
\end{align}
for a stationary state.

\section{Numerical Results}

\label{III}

To study the spontaneous condensation of dipolar $^{164}$Dy atoms  on the surface of a cylinder,
%formation of a  cylindrical-shell-shaped     dipolar BEC
the   partial differential  GP  
 equation (\ref{GP3d2}) is solved,  numerically, using C/FORTRAN programs \cite{dip} or their open-multiprocessing versions \cite{omp,ompF}, 
using  the split-time-step Crank-Nicolson
method  by imaginary- and real-time propagation \cite{crank}.
The imaginary-time propagation is employed to study the stationary states and the real-time propagation for dynamics.
  It is problematic to treat numerically the divergent $1/|{\bf R}|^3$ term in the dipolar potential (\ref{pd}) in configuration space. To avoid this problem 
 the nonlocal dipolar interaction integral in the {improved} mean-field model (\ref{GP3d2})  is calculated  in the momentum $\bf k$
 space by a Fourier transformation employing the following  convolution identity \cite{dip}
 \begin{align}\label{xcv}
 \int d{\bf r'}U_{\mathrm{dd}}({\bf R})n({\bf r'})=\int \frac{d \bf k}{(2\pi)^3}e^{-i\bf k\cdot r} \widetilde U_{\mathrm{dd}}({\bf k})  \widetilde n({\bf k}),
 \end{align}
 where density $n({\bf r})\equiv |\psi({\bf r})|^2$;   $\widetilde U_{\mathrm{dd}}({\bf k})$  and $\widetilde n({\bf k})$ 
 are the Fourier transforms of the dipolar potential and density. 
 The Fourier transformation of the dipolar potential  $\widetilde U_{\mathrm{dd}}({\bf k})$ is analytically known \cite{dip} and this enhances  the accuracy of the numerical procedure.
 The integrand in the right-hand-side of Eq. \ref{xcv} is a smooth function and can be evaluated numerically employing a fast Fourier transformation routine. In this way the problem is solved in momentum space and the solution in configuration space 
 is obtained by taking another Fourier transformation.

  For the appearance of  a cylindrical-shell-shaped state,  we need a strongly dipolar atom with 
  {$a_{\mathrm{dd}}>a$}. The system becomes repulsive for 
 { $a_{\mathrm{dd}}<a$,}
  and no such states can be formed.  Although,  $a_{\mathrm{dd}}=130.8a_0$ for  $^{164}$Dy atoms, 
we have a certain flexibility in fixing the scattering length $a$, as the scattering length can be modified by 
the Feshbach resonance technique by manipulating an external electromagnetic field.
In this study 
 we take the  scattering length $a=80a_0$, whereas its experimental estimate is    $a=(92\pm 8)a_0$  \cite{scatmes}.
 With the reduction of contact repulsion, 
 this choice of scattering length ($a=80a_0$) has the advantage of slightly increasing the net attraction, which will facilitate the formation of pronounced cylindrical-shell-shaped BEC.
For dysprosium atoms   $m(^{164}$Dy)    $\approx 164 \times 1.66054\times 10^{-27}$ kg, $\hbar =  1.0545718 \times  10^{-34}$ m$^2$ kg/s. Consequently, for $f_z =  167$ Hz, 200 Hz, and 250   Hz,  the units of length are $l=\sqrt{\hbar/m\omega_z}= 0.6075$ $\mu$m, 0.5551 $\mu$m,   and 0.4965 $\mu$m, respectively.   Incidentally, $f_z=167$ Hz is the $z$ frequency of the  trap employed in the recent experiment on 2D supersolid formation in $^{164}$Dy \cite{ex3}.  {It is
true that the frequency $f_z$ is the same in both cases, but the quasi-2D trap shape in Ref. \cite{ex3}
is
very different from the trap in the present case ($f_x,f_y = 0.75f_z$),  with much larger  trap frequencies along  the $x$ and $y$ directions,
resulting in different configurations.} The demonstration of a cylindrical-shell-shaped 
BEC is best  confirmed by a hollow in the  integrated 2D density $n_{2D}(x,y)$ defined by 
\begin{align}
n_{2D}(x,y)=\int_{-\infty}^{\infty} dz |\psi(x,y,z)|^2.
\end{align} 
 
{
\begin{figure}[t!]
\begin{center}

\includegraphics[width=\linewidth]{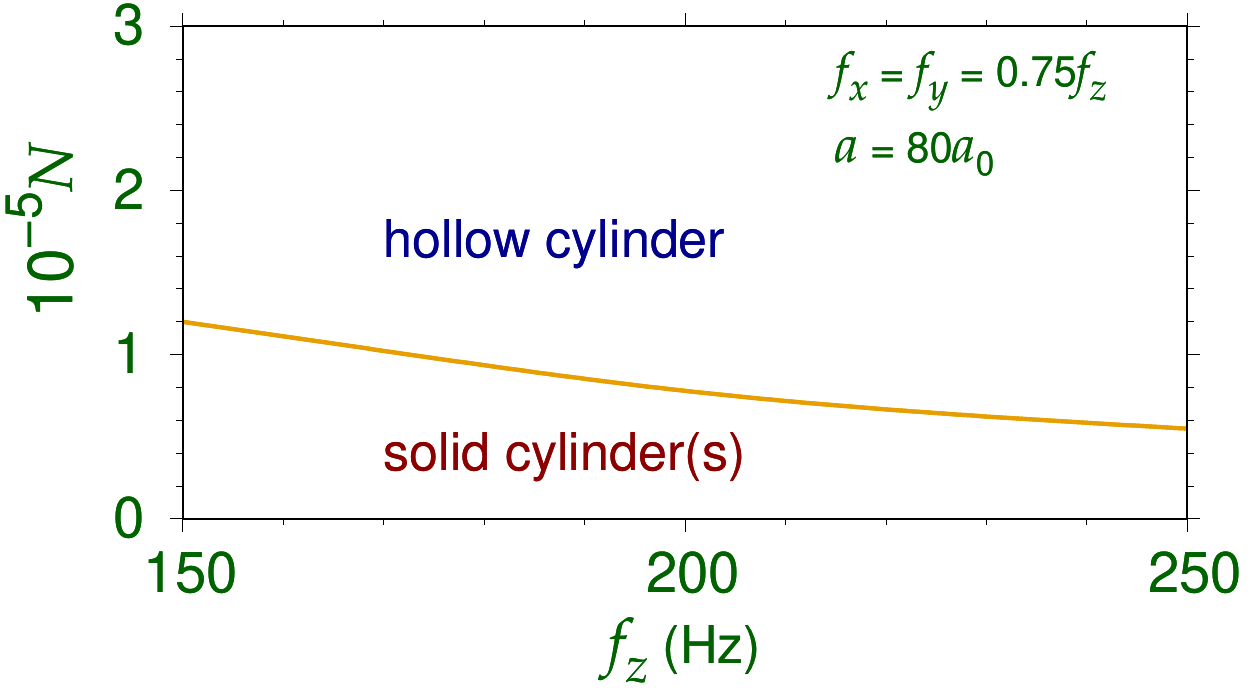}

\caption{ Phase plot of number of atoms $N$ versus trap frequency $f_z$ illustrating the formation of a 
cylindrical-shell-shaped dipolar BEC of $^{164}$Dy atoms. The region marked ``solid cylinder(s)'' include a single cylinder as well as multiple cylinders in the form of droplets.% Hollow cylindrical states with one and four holes are clearly marked.
}
\label{fig1} 
\end{center} 
\end{figure} 
}

\begin{figure}[t!]
\begin{center} 

\includegraphics[width=\linewidth]{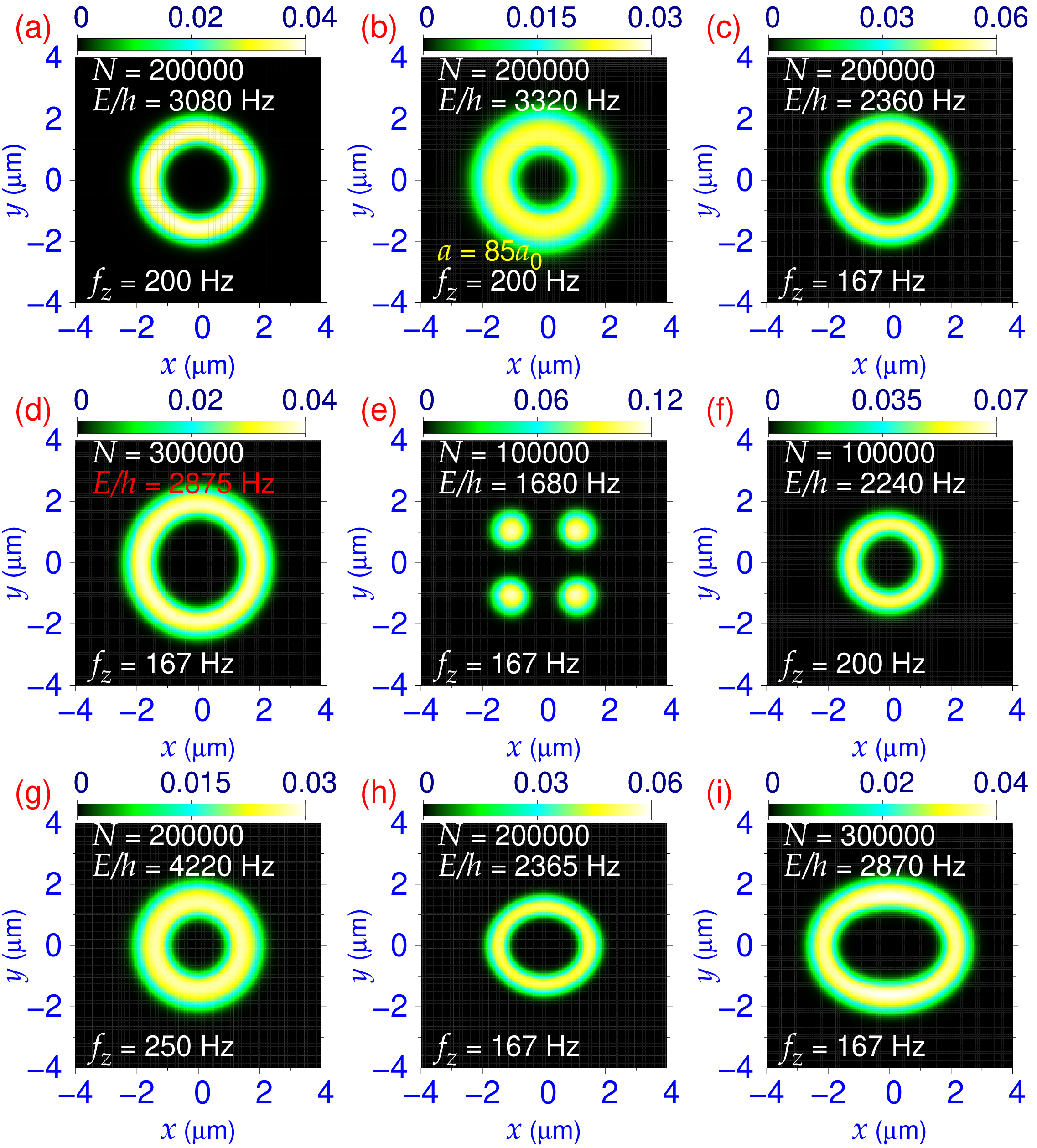}

\caption{Contour plot of dimensionless  2D density 
$n_{2D}(x,y)$ of four  cylindrical-shell-shaped states 
of $^{164}$Dy atoms ($a_{\mathrm{dd}}=130.8 a_0$) 
for (a) $N=200000,  f_z=200$ Hz, (b) $N=200000,  f_z=200$ Hz,
(c) $N=200000,  f_z=167$ Hz, and (d) {$N=300000$},  $f_z=167$ Hz; 
that of a four-droplet state for (e) $N=100000,  f_z=167$ Hz,
those of two  cylindrical-shell-shaped states for (f)  $N=100000,  f_z=200$ Hz, and 
  (g) $N=200000,  f_z=250$ Hz;    that of a cylindrical-shell-shaped state with elliptical section for  
  (h)  $N=200000,  f_z=167$ Hz, and
  (i)   $N=300000,  f_z=167$ Hz. 
In (a)-(g)  $f_x=f_y=0.75 f_z$, and in (h)-(i) $f_x=120$ Hz, and  $f_y=130$ Hz; in (b) $a=85a_0$ and in (a) and (c)-(i)  $a=80a_0$.   The energies per atom in each case appear in the $E/h$ values in Hz
displayed
in respective plots.
 %In all plots $a=80a_0$ except in (e) where $a=85a_0$.
}
\label{fig2} 
\end{center} 
\end{figure}

{The scenario  of the formation of a cylindrical-shell-shaped state in a $^{164}$Dy BEC is best illustrated by a  phase plot of the number of atoms $N$  versus trap frequency $f_z$, as obtained by imaginary-time propagation of Eq. (\ref{GP3d2}) employing a cylindrically-symmetric Gaussian initial state,
for $a=80a_0$ and $f_x=f_y=0.75f_z$ as illustrated in Fig. \ref{fig1}.  We find that, for  a fixed $f_z$, the cylindrical-shell-shaped states are obtained for the number of atoms $N$ larger than a critical value.
%The hollow cylinders have a single coaxial hole for small $f_z$ ($\lessapprox 200$ Hz). %For large $f_z$ ($\gtrapprox 210$ Hz) and $N$ ($\gtrapprox 3\times 10^5$), cylindrical states with four holes appear as ground states.
 For smaller $f_z$ ($\lessapprox 200$ Hz) and large $N$ ($\gtrapprox 3\times 10^5$) one also has the labyrinthine states \cite{pfau} (not studied here).
To obtain the  labyrinthine states, which are mostly excited states for $N<300000$,
special symmetry-broken initial states are needed in imaginary-time propagation. The cylindrically-symmetric Gaussian initial states lead to the cylindrical-shell-shaped states. 
For smaller $N$ ($\lessapprox 10^5$),  we find different states in the form of solid cylinders without any hollow part inside. These include a single solid cylinder (not shown here)
 or multiple solid cylinders aligned along the $z$ axis in the form of droplets, viz. Fig. \ref{fig2}(e). }

 A contour plot of the integrated 2D density $n_{2D}(x,y)$   of a cylindrical-shell-shaped state in a $^{164}$Dy BEC is considered next. % For $a\gtrapprox 88a_0$ the system is more repulsive and a Gaussian-type extended BEC is obtained in imaginary-time propagation, viz. Fig. \ref{fig4}(b). % Pronounced  cylindrical-shell-shaped state appears around $a\approx 80a_0$, and in this study we will employ the value $a=80a_0$.  
 In  Fig. \ref{fig2}, we illustrate (a) a cylindrical-shell-shaped state  with $N=200000, a=80a_0, f_z=200$ Hz, $f_x=f_y=0.75f_z$.   
The inner radius of the cylindrical shell in Fig. \ref{fig2}(a) is quite sharp and pronounced. 
As the scattering length $a$ is increased,  the contact repulsion increases relative to the dipolar interaction and consequently, 
the inner radius of the cylindrical  shell reduces as can be found in  the contour plot of the  cylindrical-shell-shaped state  in Fig. \ref{fig2}(b) for  $N=200000, a=85a_0, f_z=200$ Hz,  by comparing with the contour plot  in Fig. \ref{fig2}(a) for $a=80a_0$, with other parameters unchanged. 
With further increase of $a$, 
for $a\gtrapprox 88a_0$, the cylindrical-shell-shaped state loses the shell-shaped structure and 
a Gaussian-type extended BEC in the shape of a normal  solid cylinder
is obtained in imaginary-time propagation (not explicitly shown here).
On the other hand, as $f_z$ is decreased, the inner radius of the cylindrical shell increases  as can be found in the contour plot of the  cylindrical-shell-shaped state  in Fig.  \ref{fig2}(c) for   $N=200000, a=80a_0, f_z=167$ Hz,  by comparing with the contour plot in Fig.  \ref{fig2}(a) for $f_z=200$ Hz with other parameters unchanged.
If the number of atoms $N$ is increased, the inner radius of the cylindrical-shell-shaped state slightly increases as illustrated in the contour plot in Fig.  \ref{fig2}(d) for $N=300000$, $a=80a_0, f_z=167$ Hz, by comparing with the contour plot in Fig. \ref{fig2}(c) for  $N=200000$ with other parameters unchanged.  
%For much larger $N$, we could not find any cylindrical-shell-shaped state and only labyrinthine states could be found (not shown here). 
 However, if $N$ is decreased below a critical value, the cylindrical-shell-shaped states are not possible and only droplet states appear. For $f_z=167$ Hz this happens for  $N\approx 100000$  as illustrated in the contour plot of  the four-droplet state in Fig.  \ref{fig2}(e).  For $f_z=200$ Hz, this critical number is smaller and for   $N=100000$ a 
cylindrical-shell-shaped state appears, viz. Fig. \ref{fig2}(f).   For  $f_z=200$ Hz, droplet states appear for $N\lessapprox 75000$ (not displayed here).
The inner radius of the cylindrical shell reduces further at an increased frequency $f_z=250$ Hz as shown in plot in  Fig.  \ref{fig2}(g) for  $N=200000$, compare with plots in  Figs. \ref{fig2}(a) and \ref{fig2}(c) for $f_z=200$ Hz and $f_z=167$ Hz, respectively. %However, as the number of atoms increase to  $N=400000$ for $f_z=250$ Hz {a new type of cylindrical-shell-shaped state  appears as the ground state
%with four holes inside as shown in Fig. \ref{fig2}(h)  $-$ not found for smaller $f_z$ $-$, viz. plot  \ref{fig2}(d) for $f_z=167$ Hz and $N=400000$. }
%For the parameters of plot in Fig.  \ref{fig2}(h), the cylindrical-shell-shaped state with one hole becomes an excited state and does not naturally appear in imaginary-time propagation with a Gaussian initial state.
%However, as the number of atoms is reduced to $N=300000$ for $f_z=250$ Hz,
%the   cylindrical-shell-shaped state with one hole   becomes the ground state.
The circular section of the  cylindrical-shell-shaped states becomes elliptical for an asymmetric trap in the $x$-$y$ plane ($f_x\ne f_y $) as demonstrated through the contour plot of $n_{2D}(x,y)$ in 
Fig. \ref{fig2}(h) for $N=200000$, and (i) for $N=300000$,  and 
$a=80a_0,  f_z=167$ Hz, $f_x=120$ Hz,  $f_y=130$ Hz.
%Fig. \ref{fig2}(i) for $N=300000, a=80a_0,  f_z=167$ Hz, $f_x=120$ Hz,  $f_y=130$ Hz.    

\begin{figure}[t!]
\begin{center}

\includegraphics[width=\linewidth]{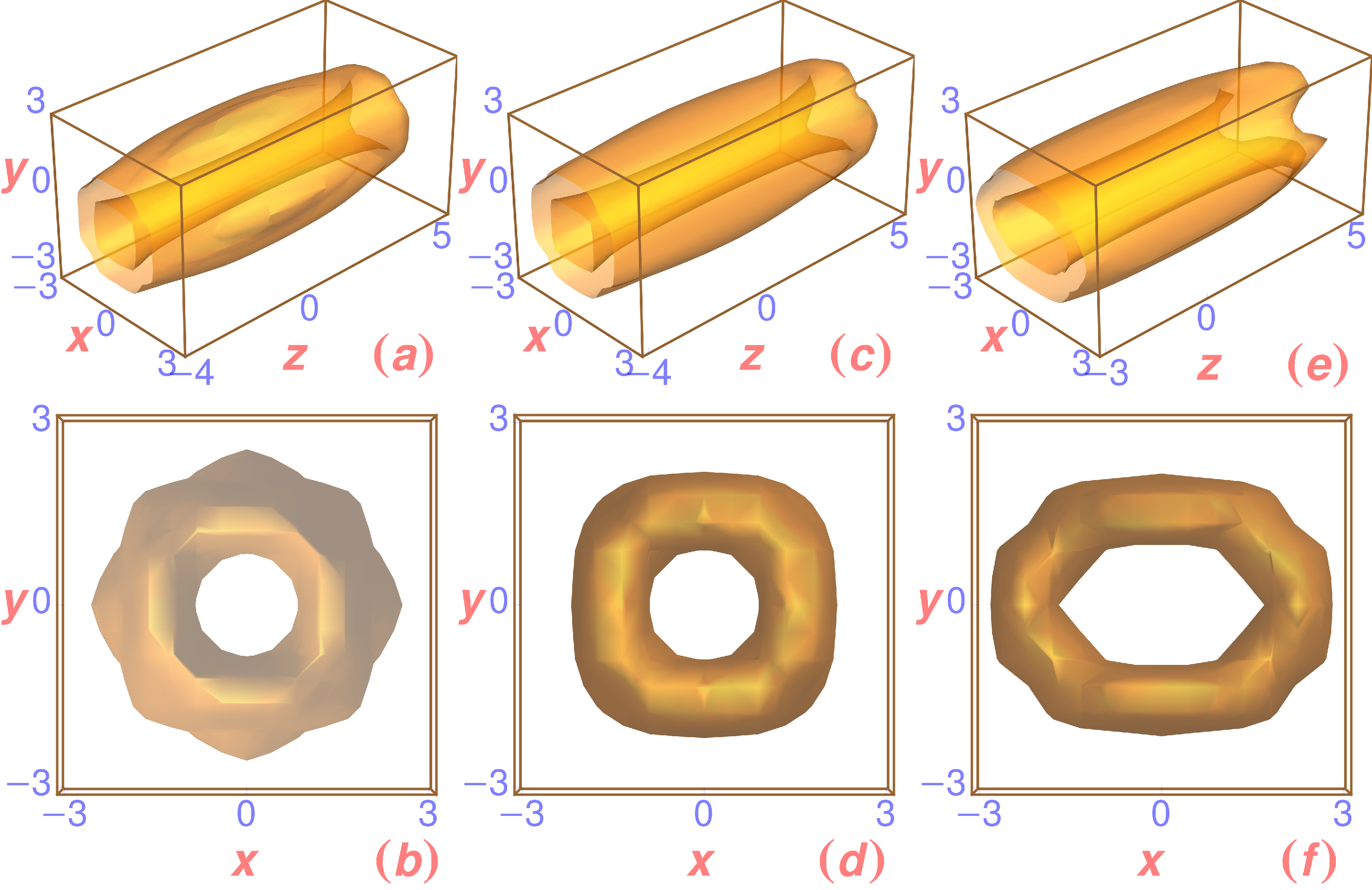}

\caption{ (a) Side view and (b) front view of isodensity plot of  $N|\psi(x,y,z)|^2$ of the 
cylindrical-shell-shaped state, viz. Fig. \ref{fig2}(a),  of $N=200000$ $^{164}$Dy atoms  in a trap with frequency $f_z=200$ Hz.
 (c) Side view and (d) front view of the same, viz. Fig. \ref{fig2}(c),  for   $f_z=167$ Hz, and $N=200000$.
{(e) Side view and (f) front view of the cylindrical-shell-shaped state with elliptic section, viz. Fig. \ref{fig2}(i),  for  $f_z=167$ Hz,  and  $N=300000$.}    The unit of  lengths $x,y$ and $z$ is $\mu$m.  
%The density on the contour is $5\%$ of the maximum density ($\gtrapprox 10^{15}$ atoms/cm$^3$). 
  In plots (a)-(d) $f_x=f_y=0.75f_z$ and $a=80a_0$ and in plots (e)-(f) $f_x=120$ Hz, $f_y=130$ Hz, and $a=80a_0$.
}
\label{fig3} 
\end{center}
\end{figure}

\begin{figure}[t!]
\begin{center} 
\includegraphics[width=.325\linewidth]{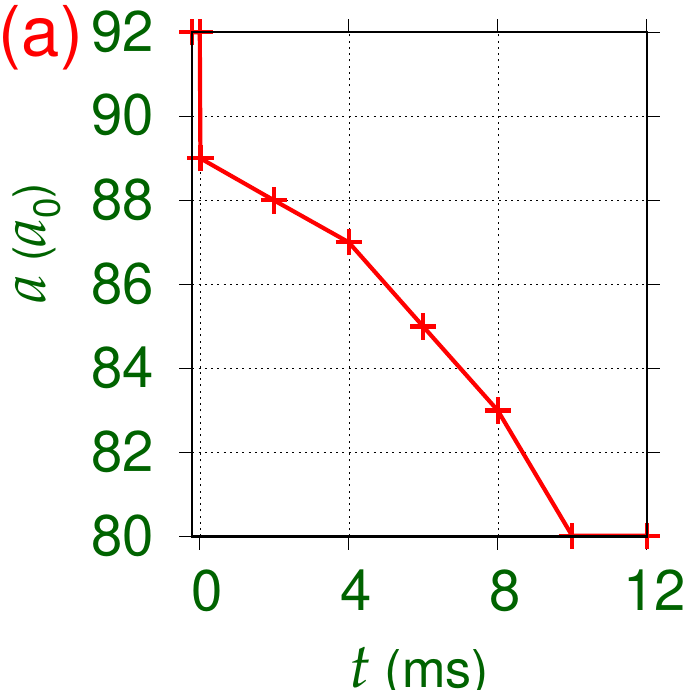}
\includegraphics[width=.325\linewidth]{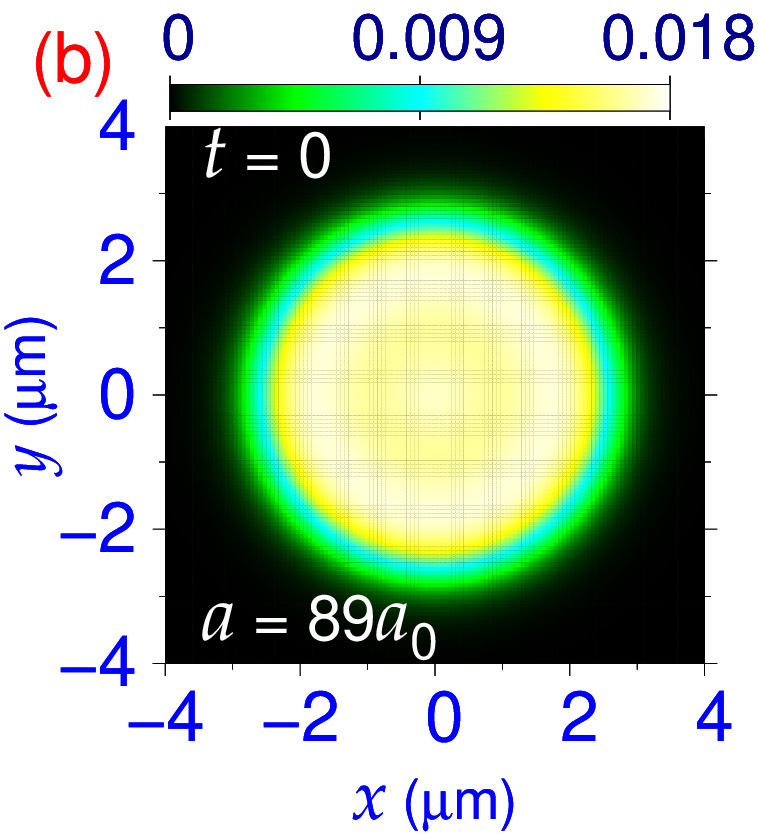}
\includegraphics[width=.325\linewidth]{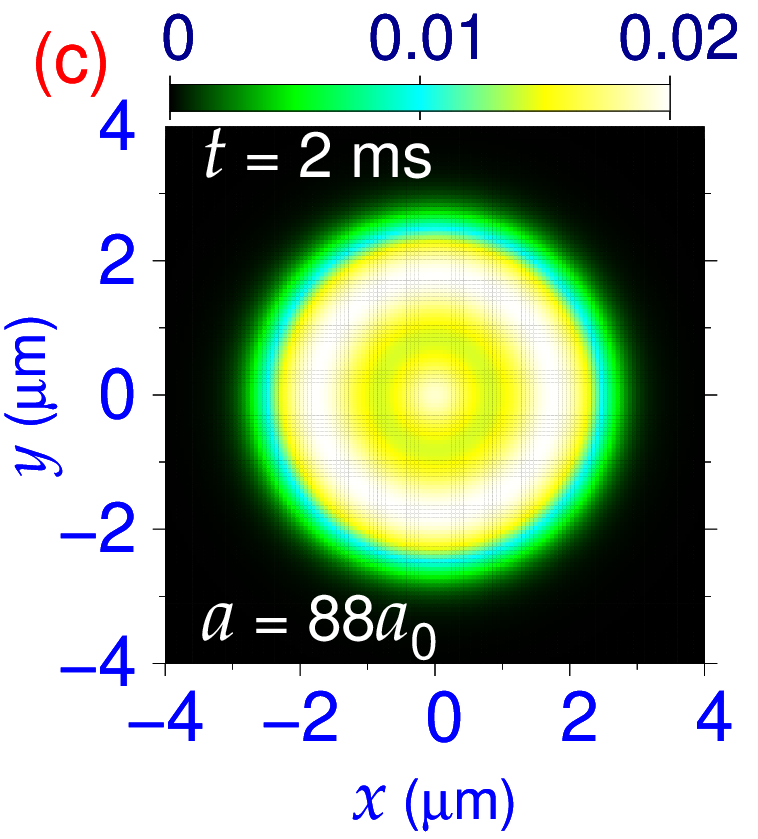}
\includegraphics[width=\linewidth]{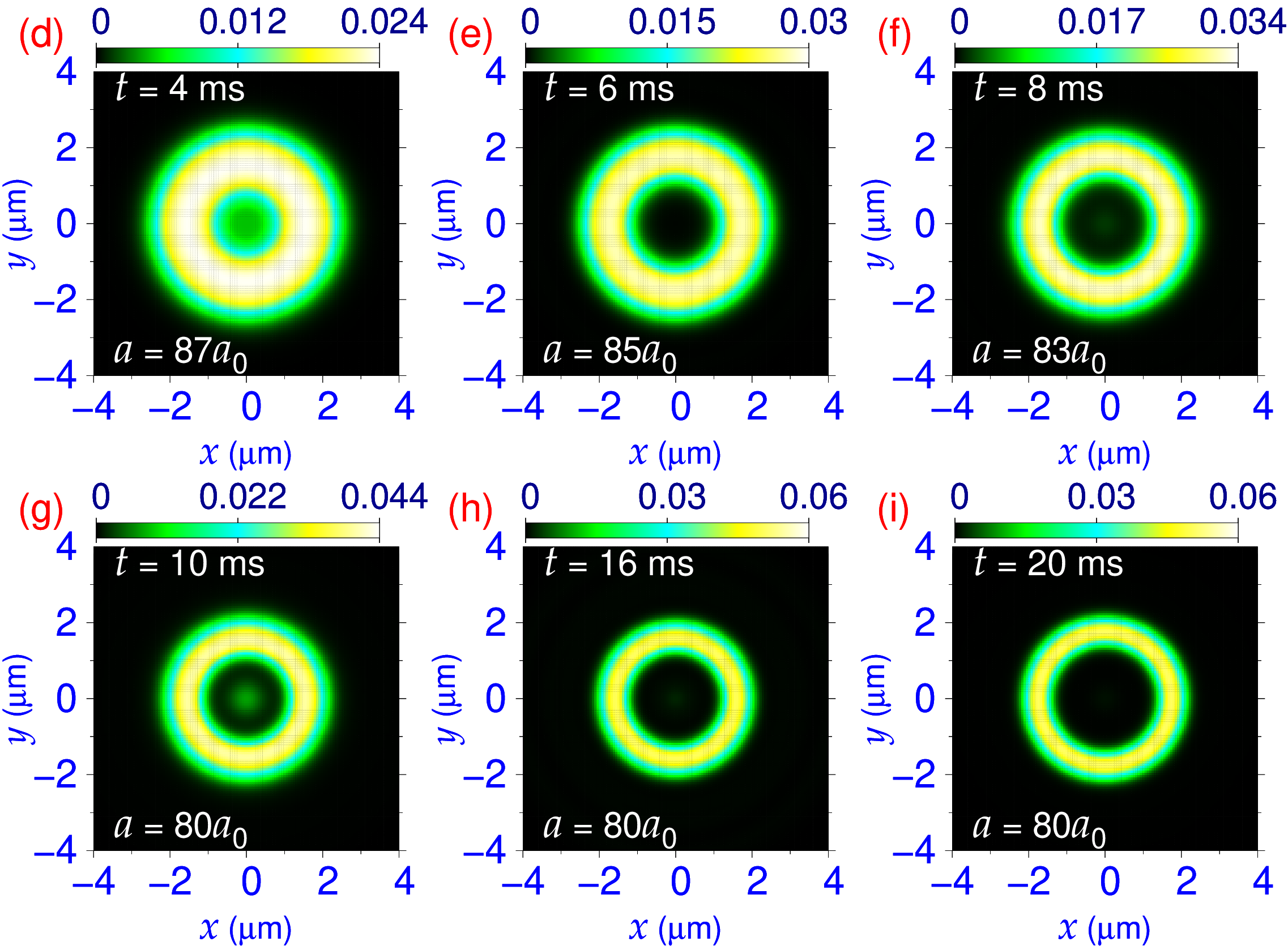}

\caption{Dynamical generation of a cylindrical-shell-shaped state  of $N=200000$ $^{164}$Dy atoms of scattering length $a=80a_0$
in a trap with frequencies $f_z=167$ Hz, $f_x=f_y=0.75f_z$ by real-time propagation initiating from a converged cylindrical state of $N=200000$ $^{164}$Dy atoms of scattering length $a=92a_0$ obtained by  imaginary-time propagation in the same trap.  (a) Scattering length versus time plot, the scattering length is changed once at an interval of 1 ms shown by the ``plus'' sign. Contour plot of 2D density $n_{2D}(x,y)$ in dimensionless units at  (b) $t=0 (a=89a_0)$ (c) $t=2$ ms $(a=88a_0)$, (d) $t=4$ ms $(a=87a_0)$, (e) $t=6$ ms $(a=85a_0)$,  (f) $t=8$ ms $(a=83a_0)$, (g) $t=10$ ms $(a=80a_0)$,   (h) $t=16$ ms $(a=80a_0)$,
(i) $t=20$ ms $(a=80a_0)$.}
\label{fig4} 
\end{center}
\end{figure}

 The extended hollow region of the cylindrical-shell-shaped state is best illustrated through the isodensity plot  of the same.  In Fig. \ref{fig3} the (a) side view and (b) front 
view of the density  $N|\psi(x,y,z)|^2$ are presented for the state with $N=200000, a=80a_0,   f_z=200$ Hz, viz. Fig. \ref{fig2}(a).  In Fig. \ref{fig3} the (c) side view and (d) front 
view of the density  $N|\psi(x,y,z)|^2$ are presented for the state with $N=200000, a=80a_0,   f_z=167$ Hz, viz. Fig. \ref{fig2}(c). Finally, in Fig. \ref{fig3} the (e) side view and (f) front 
view of the density  $N|\psi(x,y,z)|^2$ are presented for the cylindrical-shell-shaped state with elliptic section  for  $N=300000, a=80a_0,   f_z=167$ Hz, viz. Fig. \ref{fig2}(i). 
 Because of dipolar interaction the cylindrical-shell-shaped state is elongated in the polarization $z$ direction. The length of the cylindrical shells is about 8 $\mu$m.  However, in plots \ref{fig3}(a), \ref{fig3}(c) and \ref{fig3}(e) we have shown a section at $z=-4$ $\mu$m and $-3$ $\mu$m, so that the inner hollow region can be easily seen.  
  The long inner hollow region of the cylinder, explicit in plots \ref{fig3}(b), \ref{fig3}(d) and \ref{fig3}(f), is also  visible in the side view of plots \ref{fig3}(a), \ref{fig3}(c) and \ref{fig3}(f). %The four holes can be clearly seen in the section of the side view  at $z=-3$ $\mu$m in Fig. \ref{fig3}(e).
   The area of the internal hollow region has increased from plots \ref{fig3}(a) and \ref{fig3}(b) to plots \ref{fig3}(c) and \ref{fig3}(d) due to a reduction of the trap frequency from $f_z= 200$ Hz to $f_z=167$ Hz.
  
    The cylindrical-shell-shaped states
are very high density states, like the droplet states \cite{ex1,ex2}, the maximum density in the interior of the states in Figs. \ref{fig2}  is    $\gtrapprox 10^{15}$ atoms/cm$^3$, {whereas the average density inside the state is about $3\times 10^{14}$ cm$^{-3}$.  In this theoretical study we have neglected the effect
of three-body recombination loss of atoms. A matter of
concern for the experimental observation \cite{ex2,42} of a  cylindrical-shell-shaped state  is the large atom number ($N \sim 10^5$) required, where the effect of three-body
recombination loss of atoms might not be negligible \cite{th3}.
Nevertheless, a reasonably small value of the loss parameter ($= 1.25 \times 10^{-41}$  m$^6$/s) is estimated for $^{164}$Dy atoms
\cite{ex2, 29} from measurements on a thermal cloud and is
assumed to be a constant over the small range of scattering lengths near $a = 80 - 90a_0$ close to the experimental
estimate $a = 92a_0$ \cite{scatmes} and the value $a = 80a_0$ used in
this study.  The typical atomic density of
$3\times 10^{14}$ cm$^{-3}$  is  within the acceptable limit
for the formation of  cylindrical-shell-shaped states    in an experiment as established in previous experimental \cite{ex2, 29} and theoretical
\cite{th3,42} investigations.  A way  to optimize the production of dipolar BECs with more than 
$2 \times  10^5$
atoms  for the study of large-atom-number dipolar gases in the droplet and
supersolid regimes has recently appeared in the literature \cite{arch}. }

 The cylindrical shell-shaped states  can possibly be easily realized in a laboratory.  It is quite possible that, for the correct set of parameters,  these high-density  states would be  naturally formed as in the case of the  seven-droplet state on a triangular lattice \cite{ex3}. Only an experiment can test this conjecture. 
Else, they can be formed starting from a solid cylindrical state, naturally formed for  the experimental value of  scattering length $a$ $(=92a_0)$ \cite{scatmes} for $N=200000$ $^{164}$Dy atoms in a trap with frequency $f_z=167$ Hz and $f_x=f_y=0.75f_z$, as we demonstrate here by real-time propagation starting with the converged  initial state obtained by imaginary-time propagation with $a=92a_0$. During real-time propagation the scattering length $a$ is changed at intervals of 2 ms to reach $a=80a_0$ at 10 ms, via $a=89a_0$ at $t=0$,  $a=88a_0$ at $t=2$ ms, $a=87a_0$ at $t=4$ ms,
  $a=85a_0$ at $t=6$ ms,  $a=83a_0$ at $t=8$ ms and  $a=80a_0$ at $t=10$ ms. This time variation of the scattering length is displayed in Fig. $\ref{fig4}$(a). The contour plot of 2D density $n_{2D}(x,y)$ as obtained in real-time propagation is illustrated in Fig. \ref{fig4}  at times 
  (b)  $t=0$   (c) $t=2$ ms, (d) $t=4$ ms,  (e) $t=6$ ms,    (f) $t=8$ ms,   
   (g) $t=10$ ms,   (h) $t=16$ ms,    (i) $t=20$ ms.   We find that the hollow region has been created at $t=6$ ms starting from  a solid cylinder at $t=0$.    This hollow region  keeps on oscillating radially for $t> 10$ ms and eventually settles to its equilibrium value at large times.

\section{Summary}

%\label{IV} 
 
 The importance of the study of quantum states with distinct topology cannot be overemphasized \cite{top1,top2,top3}. Yet, the generation of such states requires an engineering of different control parameters as in the recent experiments with a spherical-shell-shaped \cite{BBL-space,SHL-erth,bs2}   and ring-shaped \cite{ring1,ring2} BEC.  In this paper, we have demonstrated the possibility of the formation of high-density dynamically-stable cylindrical-shell-shaped quantum states with  distinct topology in a harmonically trapped dipolar BEC of $^{164}$Dy atoms  for parameters $-$ number of atoms, trap frequencies $-$ employed in the recent experiments \cite{expt}, using an improved mean-field GP model including the LHY interaction \cite{lhy} appropriately modified for dipolar atoms \cite{qf1,qf2,qf3}. %The first type of these states have the form of a cylindrical shell, viz. Figs. \ref{fig3}(a)-(d), and the {other type has the form of a cylinder with four holes parallel to the axis of the cylinder,} viz. Figs. \ref{fig3}(e)-(f).  
 We have demonstrated by real-time simulation that, starting from an initial state with the shape of a solid cylinder at $a=92a_0$, and  by ramping down  the scattering length in a few steps to the desired value ($a=80a_0$) in a short interval of time ($\sim 10$ ms), it is possible to generate the cylindrical-shell-shaped state quickly. This ensures the possibility of observing  a cylindrical-shell-shaped  state in a laboratory in a harmonically-trapped dipolar BEC of $^{164}$Dy atoms
 in the near future.

\begin{acknowledgments}
The authors thank Dr. Vyacheslav I. Yukalov for helpful comments.
SKA acknowledges support by the CNPq (Brazil) grant 301324/2019-0.  The use of the supercomputing  cluster of the Universidad de Cartagena is acknowledged.

\end{acknowledgments}


\begin{thebibliography}{99}
% 


\bibitem{top1}M. Z. Hasan and C. L. Kane,  Rev. Mod. Phys. 82, 3045 (2010).


\bibitem{top2} X.-L. Qi and S.-C. Zhang,  Rev. Mod. Phys. 83, 1057 (2011).


\bibitem{top3} C.-K. Chiu, J. C. Y. Teo, A. P. Schnyder, and S. Ryu, 
Rev. Mod. Phys. 88, 035005 (2016).



\bibitem{curved}N. S. M\'oller, F. E. A. dos Santos, V. S. Bagnato, and A. Pelster,  New J. Phys. 22, 063059
(2020).



\bibitem{QC} A. Stern and N. H. Lindner,  Science
339, 1179 (2013).


\bibitem{vor} A. M. Turner, V. Vitelli, and D. R. Nelson, Rev. Mod. Phys. 82, 1301 (2010).



\bibitem{HH}T.-L. Ho and B.  Huang,
Phys. Rev. Lett. 115, 155304 (2015). supercyl    

\bibitem{HH1}N.-E. Guenther, P. Massignan, and A. L. Fetter,
Phys. Rev. A 96, 063608 (2017).
 
\bibitem{FQH} T. Can, M. Laskin, and P. Wiegmann,
Phys. Rev. Lett. 113, 046803 (2014).
 

\bibitem{ring1} S. Moulder, S. Beattie, R. P. Smith, N. Tammuz, and Z.
Hadzibabic,   Phys. Rev. A 86, 013629 (2012).


\bibitem{ring2} S. Eckel, J. G. Lee, F. Jendrzejewski, N. Murray, C. W.
Clark, C. J. Lobb, W. D. Phillips, M. Edwards, and G. K.
Campbell,   Nature 506, 200 (2014).


\bibitem{BBL-space}R. A. Carollo, D. C. Aveline, B. Rhyno, S. Vishveshwara, C. Lannert, J. D. Murphree, E. R. Elliott, J. R. Williams, R. J. Thompson, and  N. Lundblad, 
Nature  606, 281 (2022).




\bibitem{EO}D. C. Aveline, J. R. Williams, E. R. Elliott, C. Dutenhoffer, J. R. Kellogg, J. M. Kohel, N. E. Lay, K. Oudrhiri, R. F. Shotwell, N. Yu, and  R. J. Thompson, Nature 582, 193 (2020).

 \bibitem{lab} N. Gaaloul, M. Meister, R. Corgier, A. Pichery, P. Boegel, W. Herr, H. Ahlers, E. Charron, J.  R. Williams, R. J. Thompson, W. P. Schleich, E. M. Rasel, and  N. P. Bigelow,
 Nature Commun.  13, 7889 (2022). 
 
 \bibitem{lab2}A. Bassi, L. Cacciapuoti, S. Capozziello, S. Dell’Agnello, E. Diamanti, D. Giulini, 
 L. Iess, P. Jetzer, S. K. Joshi,
A. Landragin, C. Le Poncin-Lafitte, E. Rasel, A. Roura, C. Salomon, and H. Ulbricht,
npj Microgravity  8, 49 (2022).
 
 
\bibitem{sugg1}
O. Zobay and B. M. Garraway,
Phys. Rev. Lett. 86, 1195 (2001).

 \bibitem{sugg2}O. Zobay and  B. M. Garraway,
 Phys. Rev. A 69, 023605 (2004).
 
\bibitem{sugg3} K. Sun, K. Padavi\'c, F. Yang, S. Vishveshwara,  and C. Lannert, Phys. Rev. A 98, 013609 (2018).

\bibitem{sugg4}S. K. Adhikari,
Phys. Rev. A 85, 053631 (2012).

\bibitem{sugg5}M. Meister, A. Roura, E. M. Rasel, and W. P. Schleich, New
J. Phys. 21, 013039 (2019).


\bibitem{TS}A. Tononi and L. Salasnich, Phys. Rev. Lett. 123, 160403 (2019).

\bibitem{SHL-erth}F. Jia, Z. Huang, L. Qiu, R. Zhou, Y. Yan, and D. Wang,
Phys. Rev. Lett. 129, 243402 (2022).
 
 
\bibitem{Ho}T.-L. Ho and V. B. Shenoy, Phys. Rev. Lett. 77, 3276
(1996).
 


\bibitem{Pu}H. Pu and N. P. Bigelow,  Phys. Rev. Lett. 80, 1130 (1998).


\bibitem{bin1}M. Trippenbach, K. G\'oral, K. Rzazewski, B. Malomed,
and Y. B. Band,  J. Phys. B 33, 4017 (2000).
\bibitem{bin2}K. L. Lee, N. B. Jorgensen, I. Kang Liu, L. Wacker, J. J.
Arlt, and N. P. Proukakis,  Phys.
Rev. A 94, 013602 (2016).

\bibitem{bin3} A. Wolf, P. Boegel, M. Meister, A. Bala\v z, N. Gaaloul,
and M. A. Efremov,  Phys. Rev. A 106,
013309 (2022).

 


\bibitem{bs2} Y. Guo, E. Mercado Gutierrez, D. Rey, T. Badr, A. Perrin, L. Longchambon, V. S.  Bagnato,   H. Perrin,  and R.
Dubessy,  
New J. Phys. 24, 093040 (2022).

 


\bibitem{lhy}T. D. Lee, K. Huang, and C. N. Yang, Phys. Rev. 106, 1135
(1957).

\bibitem{qf1}A. R. P.  Lima and  A. Pelster,   Phys. Rev. A 84, 041604(R) (2011).


 \bibitem{qf3}A. R. P. Lima and A. Pelster,
Phys. Rev. A 86, 063609 (2012).

\bibitem{qf2}{R. Sch\"utzhold, M. Uhlmann, Y. Xu, and U. R. Fischer, Int. J. Mod. Phys. B 20,  3555 (2006).}
 
\bibitem{ex1} H. Kadau, M. Schmitt, M. Wenzel, C. Wink, T. Maier, I.
Ferrier-Barbut, and T. Pfau, Nature  530, 194
(2016).

\bibitem{ex2}M. Schmitt, M. Wenzel, F. B\"ottcher, I. Ferrier-Barbut,
and T. Pfau, Nature 539, 259 (2016).


\bibitem{santos} F. W\"achtler and L. Santos,
Phys. Rev. A 93, 061603(R) (2016).



\bibitem{ex3}M. A. Norcia, C. Politi, L. Klaus, E. Poli, M. Sohmen,
M. J. Mark, R. Bisset, L. Santos, and F. Ferlaino, Nature
 596, 357 (2021).


\bibitem{th4} Luis E. Young-S. and S. K. Adhikari, Phys. Rev. A 105,
033311 (2022).

\bibitem{pfau} J. Hertkorn, J.-N. Schmidt, M. Guo, F. B\"ottcher, K. S. H. Ng, S. D. Graham, P. Uerlings, T. Langen, M. Zwierlein, and T. Pfau,
Phys. Rev. Research 3, 033125 (2021).

\bibitem{science}A. J. Dickstein, S. Erramilli, R. E. Goldstein, D. P. Jackson,
and S. A. Langer,  Science 261, 1012 (1993).

\bibitem{th1}E. Poli, T. Bland, C. Politi, L. Klaus, M. A. Norcia, F.
Ferlaino, R. N. Bisset, and L. Santos, Phys. Rev. A 104,
063307 (2021).


\bibitem{th2}D. Baillie and P. B. Blakie, Phys. Rev. Lett. 121, 195301
(2018).



\bibitem{th3}Y.-C. Zhang, T. Pohl, and F. Maucher, Phys. Rev. A 104,
013310 (2021).


\bibitem{th5} L. E. Young-S. and S. K. Adhikari, Eur. Phys. J. Plus
137, 1153 (2022).

\bibitem{expt}L. Chomaz, I. Ferrier-Barbut, F. Ferlaino, B. Laburthe-Tolra, B. L. Lev, and T. Pfau,
Rep. Prog. Phys.   86,  026401 (2023).
 

 \bibitem{arch}M. Krstaji\'c, P. Juh\'asz, J. Ku\v cera, L. R. Hofer, G. Lamb, A. L. Marchant, R. P. Smith, arXiv:2307.01245. 
 


\bibitem{scatmes} Y. Tang, A. Sykes, N. Q. Burdick, J. L. Bohn, and B. L. Lev,
Phys. Rev. A 92, 022703 (2015).


\bibitem{dipbec}T Lahaye, C Menotti, L Santos, M Lewenstein, and T
Pfau, Rep. Prog. Phys. 72, 126401 (2009).


\bibitem{blakie}R. N. Bisset, R. M. Wilson, D. Baillie, and P. B. Blakie,
Phys. Rev. A 94, 033619 (2016).

\bibitem{dip} R. Kishor Kumar, L. E. Young-S., D. Vudragovi\'c, A. Bala\v{z}, P. Muruganandam, and S. K. Adhikari, Comput. Phys. Commun. { 195}, 117 (2015).



\bibitem{yuka}V. I. Yukalov, Laser Phys.  28, 053001 (2018).

 
\bibitem{young} Luis E. Young-S. and S. K. Adhikari,
Commun. Nonlin. Sci. Numer. Simul. 115, 106792 (2022).

\bibitem{ompF} L. E. Young-S., P. Muruganandam, A. Bala\v z, and S. K. Adhikari,
Comput. Phys. Commun.  286, 108669 (2023).

\bibitem{omp}V. Lon\v car, L. E. Young-S., S. \v Skrbi\'c, P. Muruganandam, S. K. Adhikari, and A. Bala\v z, Comput. Phys. Commun. { 209}, 190 (2016).





\bibitem{crank}P. Muruganandam and S. K. Adhikari, Comput. Phys. Commun. 180, 1888       (2009).




\bibitem{42}Y.-C. Zhang, F. Maucher, and T. Pohl, Phys. Rev. Lett.
123, 015301 (2019).
 
\bibitem{29} I. Ferrier-Barbut, H. Kadau, M. Schmitt, M. Wenzel, and
T. Pfau, Phys. Rev. Lett. 116, 215301 (2016).
 
 
\end{thebibliography}
\end{document}